\documentstyle[times,psfig,color,mathrsfs]{mn}

\newif\ifAMStwofonts
\AMStwofontstrue

\ifoldfss

  \newcommand{\tbd}[1] {\textbf{\color{red}{To be done later}}}
  \ifCUPmtlplainloaded \else
    \NewTextAlphabet{textbfit} {cmbxti10} {}
    \NewTextAlphabet{textbfss} {cmssbx10} {}
    \NewMathAlphabet{mathbfit} {cmbxti10} {} % for math mode
    \NewMathAlphabet{mathbfss} {cmssbx10} {} %  "   "    "
  \fi
  \ifAMStwofonts
    \ifCUPmtlplainloaded \else
      \NewSymbolFont{upmath} {eurm10}
      \NewSymbolFont{AMSa} {msam10}
      \NewMathSymbol{\upi}     {0}{upmath}{19}
      \NewMathSymbol{\umu}     {0}{upmath}{16}
      \NewMathSymbol{\upartial}{0}{upmath}{40}
      \NewMathSymbol{\leqslant}{3}{AMSa}{36}
      \NewMathSymbol{\geqslant}{3}{AMSa}{3E}

      \let\leq=\leqslant 
       
    \fi
  \fi
\fi % End of OFSS

\ifnfssone
  \newmathalphabet{\mathit}
  \addtoversion{normal}{\mathit}{cmr}{m}{it}
  \addtoversion{bold}{\mathit}{cmr}{bx}{it}

  \newcommand{\tbd}[1] {\textbf{\color{red}{To be done later}}}
  \newmathalphabet{\mathbfit} % math mode version of \textbfit{..}
  \addtoversion{normal}{\mathbfit}{cmr}{bx}{it}
  \addtoversion{bold}{\mathbfit}{cmr}{bx}{it}
  \newmathalphabet{\mathbfss} % math mode version of \textbfss{..}
  \addtoversion{normal}{\mathbfss}{cmss}{bx}{n}
  \addtoversion{bold}{\mathbfss}{cmss}{bx}{n}
  \ifAMStwofonts
    \ifCUPmtlplainloaded \else
      \UseAMStwoboldmath
      \makeatletter
      \new@mathgroup\upmath@group
      \define@mathgroup\mv@normal\upmath@group{eur}{m}{n}
      \define@mathgroup\mv@bold\upmath@group{eur}{b}{n}
      \edef\UPM{\hexnumber\upmath@group}
      \new@mathgroup\amsa@group
      \define@mathgroup\mv@normal\amsa@group{msa}{m}{n}
      \define@mathgroup\mv@bold\amsa@group{msa}{m}{n}
      \edef\AMSa{\hexnumber\amsa@group}
      \makeatother
      \mathchardef\upi="0\UPM19
      \mathchardef\umu="0\UPM16
      \mathchardef\upartial="0\UPM40
      \mathchardef\leqslant="3\AMSa36
      \mathchardef\geqslant="3\AMSa3E

      \let\leq=\leqslant 

    \fi
  \fi
\fi % End of NFSS release 1

\ifnfsstwo

  \newcommand{\tbd}[1] {\textbf{\color{red}{To be done later}}}
  \DeclareMathAlphabet{\mathbfit}{OT1}{cmr}{bx}{it}
  \SetMathAlphabet\mathbfit{bold}{OT1}{cmr}{bx}{it}
  \DeclareMathAlphabet{\mathbfss}{OT1}{cmss}{bx}{n}
  \SetMathAlphabet\mathbfss{bold}{OT1}{cmss}{bx}{n}
  \ifAMStwofonts
    \ifCUPmtlplainloaded \else
      \DeclareSymbolFont{UPM}{U}{eur}{m}{n}
      \SetSymbolFont{UPM}{bold}{U}{eur}{b}{n}
      \DeclareSymbolFont{AMSa}{U}{msa}{m}{n}
      \DeclareMathSymbol{\upi}{0}{UPM}{"19}
      \DeclareMathSymbol{\umu}{0}{UPM}{"16}
      \DeclareMathSymbol{\upartial}{0}{UPM}{"40}
      \DeclareMathSymbol{\leqslant}{3}{AMSa}{"36}
      \DeclareMathSymbol{\geqslant}{3}{AMSa}{"3E}

      \let\leq=\leqslant 

    \fi
  \fi
\fi % End of NFSS release 2

\ifCUPmtlplainloaded \else
  \ifAMStwofonts \else % If no AMS fonts
    \def\upi{\pi}
    \def\umu{\mu}
    \def\upartial{\partial}
  \fi
\fi

\title[The Chemical Evolution of the Milky Way: the Three Infall Model]{The Chemical Evolution of the Milky Way: the Three Infall Model}
\author[A.~Micali, F.~Matteucci and D.~Romano]{A.~Micali,$^{1}$\thanks{E-mail: micali@oats.inaf.it} F.~Matteucci$^{1,2}$\thanks{E-mail: matteucc@oats.inaf.it} and D.~Romano $^{3}$\thanks{E-mail: donatella.romano@oabo.inaf.it}\\
$^{1}$Dipartimento di Fisica, Sezione di Astronomia, Universit\`a di Trieste, via G.~B. Tiepolo 11, I-34143 Trieste, Italy\\
$^{2}$INAF, Osservatorio Astronomico di Trieste, via G.~B. Tiepolo 11, I-34143 Trieste, Italy\\
$^{3}$INAF, Osservatorio Astronomico di Bologna, via Ranzani 1, I-40127 Bologna, Italy}
  
\begin{document}

\date{Accepted 2013 September 4. Received 2013 September 4; in original form 2013 June 24}

\pagerange{\pageref{firstpage}--\pageref{lastpage}} \pubyear{2013}

\maketitle

\label{firstpage}

\begin{abstract}
We present a new chemical evolution model for the Galaxy that assumes three main infall episodes of primordial gas for the formation of halo, thick and thin disk, respectively. We compare our results with selected data taking into account NLTE effects. 
The most important parameters of the model are (i) the timescale for gas accretion, (ii) the efficiency of star formation and (iii) a threshold in the gas density for the star formation process, for each Galactic component. 
We find that, in order to best fit the features of the solar neighbourhood, the halo and thick disk must form on short timescales ($\sim$0.2 and $\sim$1.25~Gyr, respectively), while a longer timescale is required for the thin-disk formation. The efficiency of star formation must be maximum (10~Gyr$^{-1}$) during the thick-disk phase and minimum (1~Gyr$^{-1}$) during the thin-disk formation. Also the threshold gas density for star formation is suggested to be different in the three Galactic components. 
Our main conclusion is that in the framework of our model an independent episode of accretion of extragalactic gas, which gives rise to a burst of star formation, is fundamental to explain the formation of the thick disk. 
We discuss our results in comparison to previous studies and in the framework of modern galaxy formation theories.
\end{abstract}


\section{Introduction}

The study of the chemical evolution of galaxies allows us to understand how the chemical abundances of the most common chemical elements evolve in space and time in the interstellar medium (ISM). 
In particular, the study of our Galaxy from a chemical point of view represents a powerful tool to investigate its formation and  evolution. 
Generally, a good agreement between model predictions and observations is obtained by assuming that the disk formed by infall of primordial gas, as originally suggested in the works of Larson (1972) and Lynden-Bell (1975), followed by many others. 
In the past years, a great deal of theoretical work has appeared concerning the chemical evolution of the Milky Way, such as purely  chemical evolution models (Chiosi 1980; Matteucci \& Greggio 1986; Tosi 1988; Matteucci \& Fran\c{c}ois 1989; Pagel 1989; Matteucci \& Fran\c{c}ois 1992; Carigi 1994; Ferrini et al. 1994; Pardi \& Ferrini 1994; Giovagnoli \& Tosi 1995; Pardi, Ferrini \& Matteucci 1995; Prantzos \& Aubert 1995; Timmes, Woosley \& Weaver 1995; Chiappini, Matteucci \& Gratton 1997, hereafter CMG97; Boissier \& Prantzos 1999; Chang et al. 1999; Chiappini et al. 1999; Portinari \& Chiosi 1999, 2000; Goswami \& Prantzos 2000; Chiappini, Matteucci \& Romano 2001; Renda et al. 2005; Cescutti et al. 2006; Romano et al. 2010), viscous models (Lacey \& Fall 1985; Sommer-Larsen \& Yoshii 1989, 1990; Tsujimoto et al. 1995), inhomogeneous models (Malinie et al. 1993) and chemodynamical models (Burkert, Truran \& Hensler 1992; Samland \& Hensler 1996). 
In particular, CMG97, taking into account several observational constraints such as the metallicity distribution function (MDF) of long-lived stars, the solar abundances and the Type Ia and Type II supernova rates at the present time, have built the Two Infall Model (2IM), suggesting a scenario in which the Galaxy forms as a result of two main infall episodes. During the first episode the halo and the thick disk form and the gas lost by the halo rapidly accumulates in the center with the consequent formation of the bulge. During the second episode, a much slower infall of gas gives rise to the thin disk with the gas accumulating in the inner regions faster than in the outer regions. This mechanism for disk formation is known as {\itshape inside-out} scenario (Larson 1976; Matteucci \& Fran\c{c}ois 1989). The chemical composition of the gas in the two infall episodes was assumed to be primordial. 
In this scenario, the formation of the halo-thick disk and thin disk are almost completely disentangled although some halo gas falls into the thin disk. 
This model suggested that to reproduce the chemical evolution of the Galaxy the timescale for halo-thick disk formation should be $\sim$0.8 Gyr, and the timescale for thin-disk formation $\sim$6--8 Gyr at the solar position. 
It is worth noting that in the 2IM, the formation of the thick disk was included in that of the halo, therefore it could not give a complete description of this Galactic zone. 
Moreover, in the 2IM a threshold in the star formation rate (SFR) is present and it is equal to 4~M$_{\odot}$ pc$^{-2}$ in the halo-thick disk phase, and  7~M$_{\odot}$ pc$^{-2}$ in the thin-disk phase. It is worth recalling that a surface gas density threshold for star formation has been observed in a variety of objects, including normal spirals, starburst galaxies and low surface brightness galaxies (Kennicutt 1989, 1998; van der Hulst et al. 1993). 
The gas density thresholds adopted by CMG97 produced results in very good agreement with the majority of observational constraints in the solar neighborhood and in the whole disk (Chiappini et al. 2001). In particular, one of the main effects of the threshold was that it naturally produced a hiatus in the SFR between the end of the halo-thick disk phase and the beginning of the thin-disk phase. Such a hiatus seems to be observed both in the plot of [Fe/O] vs [O/H] (CMG97; Gratton et al. 1996, 2000) and in the plot of [Fe/Mg] vs [Mg/H] (Fuhrmann 1998; Gratton et al. 2000). 
In the last years, a great deal of observational data on abundances in thick-disk stars has appeared. The most recent observational data show that, for [Fe/H]$<0$, the thin- and thick-disk abundance trends are different (e.g. Bensby et al. 2005), especially for the $\alpha$-elements, namely those elements that form through $\alpha$-capture processes (for instance Mg, Si, S and Ca). The abundance trends we see are also well defined and with small overall scatter. This is indicative of the fact that both the thin and thick disks formed in two independent processes (Bensby et al. 2005; Bensby \& Feltzing 2006). 
The aim of this paper is to reproduce the characteristics of the thick-disk stars including the MDF and the abundance patterns. 
To do so, we study the evolution of the thick disk starting from the comparison between the 2IM with the observational data and suggesting a new model which treats the thick disk independently from the halo and the thin disk. 
The chemical evolution of the thick disk has been studied by means of both classical chemical evolution models and models based on the hierarchical picture of galaxy formation. In particular, in the classic model of Pardi et al. (1995) the thick disk forms out of gas shed by the halo and the thin disk forms out of gas shed by the thick disk. However, Pardi et al. (1995) were not able to well reproduce at the same time the MDFs of the halo and thick disk. In order to do that, one needs to assume independent episodes of formation for these two components. On the other hand, in Chang et al. (1999) it is assumed that the formation of both thick and thin disk occurs in two main accretion episodes with both infall rates being gaussian. Chiappini (2008) suggested that the formation of the thick disk could be described by a separate and independent infall episode, relative to the formation of the halo and the thin disk. 
All of the previous models assumed that the thick disk formed by smooth gas accretion, while other models proposed the formation of the thick disk by building blocks made of gas (Sommer-Larsen et al. 2003; Brook et al. 2004). Finally, hierarchical models of galaxy formation have suggested that the thick disk forms by accretion of stars originating in Milky Way satellites. 
We develop the  idea of an infall episode forming the thick disk in between the two episodes relative to the halo and thin-disk formation, and provide a new description of the formation and chemical evolution of the Milky Way. In particular, we present a new chemical evolution model, that assumes three main episodes of infall of primordial gas, and we call it Three Infall Model (3IM). The first episode is responsible for the halo formation; during the second, the thick disk is produced. Finally, a much slower third infall of primordial gas gives rise to the thin disk. 
The 3IM will be tested relative to a maximum number of the most recent observational constraints. 
In particular, the 3IM will allow us to predict: 
\begin{enumerate}
\item the MDF of the thick disk, 
\item the MDF of the thin disk, 
\item the present Type Ia and Type II supernova rates,
\item the behavior of oxygen and $\alpha$-element (magnesium, silicon and calcium) abundances, as a function of [Fe/H] for the halo, thick and thin disk. 
\end{enumerate}

The layout of the paper is as follows. In Section~2 we describe the adopted data set. The 2IM and the 3IM, with the relative nucleosynthesis prescriptions, are introduced in Sections~3 and 4, respectively. In Section~5 we present the results of the 3IM, obtained through a comparison with the most recent observational constraints. Finally, in Section~6, we discuss the results and draw our conclusions. 

\section[]{Observational Data} \label{obdata}

In order to obtain a good comparison between model predictions and observational data, we need a data set as homogeneous as possible. This is especially important for the Galactic disk, which must be sectioned in thick and thin disk. As this separation is still difficult (see e.g. Steinmetz 2012, and references therein), we must be sure that the selection method and the adopted model of atmosphere are quite similar. Moreover, it is preferable to employ only data in which NLTE corrections are considered. 
In particular, the data we use are from:

\begin{itemize}
\item Gratton et al. (2003), Cayrel et al. (2004), Akerman et al. (2004),  Mashonkina et al. (2007) and Shi et al. (2009) (for the available elements) for halo stars;
\item Gratton et al. (2003), Ructhi et al. (2011), Bensby et al. (2005), Bensby \& Feltzing (2006), Reddy et al. (2006) (except oxygen as NLTE corrections are not provided for this element) and Ramya et al. (2012) for thick-disk stars;
\item Gratton et al. (2003), Bensby et al. (2005), Bensby \& Feltzing (2006) and Reddy et al. (2003) for thin-disk stars.
\end{itemize}

As regards to the halo, the principal advantage of using Mashonkina et al. (2007) and Shi et al. (2009) is that both studies look at the same stars, use exactly the same models, spectra and atomic data, although the sample size is rather limited. 
For the thick disk, the data sets are quite good. In Ructhi et al. (2011) there are some systematical errors in the $T_{\mathrm{eff}}$, although they are not large and affect mostly the abundance scatter at a given [Fe/H] value. 

For the thin disk, Reddy et al. (2003, 2006) adopt an unbiased $T_{\mathrm{eff}}$ scale, but there could be a possible effect on [Fe/H] due to un-accounted NLTE. Also, their Si abundances are clearly under-estimated due to the neglection of NLTE, because Si\,{\sevensize I} lines were used. 
Moreover, we have preferred not to use any data set older than 2000, mostly because they used outdated spectral analysis codes, atomic data and model atmospheres, or low-resolution  spectra. For example, it is known that low-resolution underestimates abundances in some cases. Also, some key atomic data for Fe\,{\sevensize II} transitions were revised only in 2009, so most of the older studies used not precise data. 
It is worth recalling that all the cited works indicate that halo stars generally  show overabundances of $\alpha$-elements relative to Fe. Thick-disk stars also show $\alpha$-enhancements, lower or comparable to those of halo stars. On the other hand, the thin-disk stars show a continuos decrease of the [$\alpha$/Fe] ratios with [Fe/H]. The data are shown in Figs. 5, 6 and 7 together with the model results.

\section[]{The Two Infall Model} \label{two_infall_model}

This model, which has been quite successfull in reproducing the main features of the Milky Way, assumes two main infall episodes: during the first, the halo and the thick disk are formed, and during the second, delayed relative to the first one, the thin disk forms. The origin of the gas in the infall episodes is extragalactic and its composition is primordial. 
The Galactic disk is approximated by several independent rings, 2~kpc wide, without exchange of matter between them, and the basic equations for the evolution of chemical elements in the gas are:

\begin{displaymath}
\frac{d\sigma_i}{dt} =  -\psi(t)X_i(t)+\int_{M_L}^{M_{Bm}}\psi(t-\tau_m)Q_{mi}(t-\tau_m)\phi(m)dm
\end{displaymath}
\begin{displaymath}
+A_B\int_{M_{Bm}}^{M_{BM}}\phi(m)\Big[\int_{\mu_{Bmin}}^{0.5}f(\mu_B)\psi(t-\tau_{m_2})Q_{mi}(t-\tau_{m_2})d\mu_B\Big]dm 
\end{displaymath}
\begin{displaymath}
+(1-A_B)\int_{M_{Bm}}^{M_{BM}}\psi(t-\tau_m)Q_{mi}(t-\tau_m)\phi(m)dm 
\end{displaymath}
\begin{displaymath}
+\int_{M_{BM}}^{M_U}\psi(t-\tau_m)Q_{mi}(t-\tau_m)\phi(m)dm 
\end{displaymath}
\begin{equation}
+\frac{d\sigma_{i,\,inf}}{dt}
\label{eq_modello}
\end{equation}

For an exhaustive explanation of this equation see CMG97. Here we will recall the main parameters: the first term on the right-hand side of equation (1) represents the effect of the SFR, while all the integrals are the rates of restitution of processed and unprocessed material from dying stars in different mass ranges. The second integral, in particular, represents the contribution of the SNe Ia and $A_B$ is the fraction of binary systems with the right characteristics to give rise to SN Ia explosions. It is assumed $A_B=$~0.04. The underlying progenitor model for Type Ia SNe is the single degenerate scenario where a white dwarf accretes material from a younger companion and explodes when reaches the Chandrasekhar mass limit ($M_{C} \sim$ 1.4~M$_{\odot}$. The function $f(\mu)$ represents the distribution function of the ratio $\mu={M_1 \over (M_1+M_2)}$, with $M_1$ and $M_2$ being the primary and secondary mass of the binary system, respectively, and $M_B=M_1 + M_2$ being the total mass of the binary system. 
The mass $M_{Bm}$ represents the minimum total mass of a binary system giving rise to a SN Ia and it is assumed to be 3~M$_{\odot}$, while $M_{BM}=$~16 M$_{\odot}$ is the maximum total mass (8+8~M$_{\odot}$) (see Greggio \& Renzini 1983). 
The quantity $Q_{mi}$ is the term containing the stellar yields: in particular, it is the mass fraction of the element $i$ which is restored into the ISM by a star of mass $m$.

The last term in the equation is the gas accretion rate; in particular, the rate of gas accretion in each shell is assumed to be:

\begin{equation}
\frac{d\sigma_i(r,t)_{inf}}{dt}= A(r)(X_i)_{inf}e^{-\frac{t}{\tau_H}} \\
+ B(r)(X_i)_{inf}e^{-\frac{t-t_{max}}{\tau_D}}
\label{rate_2im}
\end{equation}
with $i$ representing a generic chemical element, and $\sigma_i(r,t)_{inf}$ is the surface mass density of the infalling material in the form of the element {\itshape i}, $(X_i)_{inf}$ gives the composition of the infalling gas, which we assumed to be primordial, and $t_{max}$ is the time of maximum accretion onto the disk; it is set equal to 1 Gyr and roughly corresponds to the end of the halo-thick disk phase. $\tau_H$ and $\tau_D$ are the timescales for mass accretion in the halo-thick disk and thin-disk components, respectively. These timescales are free parameters of the model, and they are constrained mainly by the comparison with the observed metallicity distribution of long-lived stars in the solar vicinity. 
In particular, $\tau_H=$~0.8 Gyr and, according to the inside-out scenario, $\tau_D(r)=$~(1.03~$r$/kpc$-$1.27)~Gyr (see Romano et al. 2000). 
This equation was constructed in order to obtain a timescale for the bulge formation ($r \leq2$ kpc) of 1 Gyr, in agreement with the results of Matteucci \& Brocato (1990), and a timescale of formation of 7~Gyr at the solar neighborhood, which best reproduces the G-dwarf metallicity distribution.

The quantities $A(r)$ and $B(r)$ are derived from the condition of reproducing the current total surface mass density distribution along the thin disk. The current total surface mass distribution is taken from Kuijken \& Gilmore (1991).

For the SFR, it is adopted an expression which has the same functional form for both the halo-thick and the thin-disk phases (CMG97). This expression, which is essentially a Schmidt (1959) law for star formation, contains a dependence also on the total surface mass density, as suggested by Talbot \& Arnett (1975) and Chiosi (1980):

\begin{equation}
\psi(r,t)=\tilde{\nu}\Big[\frac{\sigma(r,t)}{{\sigma}({r_{\odot}},t)}\Big]^{2(k-1)}
\Big[\frac{\sigma(r,t_G)}{{\sigma}(r,t)}\Big]^{k-1}
G^k(r,t)
\end{equation}
where $\tilde{\nu}$ is the efficiency of the star formation process expressed in units of Gyr$^{-1}$, $\sigma(r_{\odot},t)$ is the total surface mass density at the solar radius $r_{\odot}$ and given time {\itshape t}, $t_G$ is the Galactic lifetime (13.7~Gyr; Bennett et al. 2012) and $G^k(r,t)$ is the surface gas density normalized to the present-time total surface mass density at each Galactocentric radius. This particular formulation of the SFR has been chosen for the sake of continuity with our previous papers. However, we run models also with a simple Kennicutt law, SFR~$\propto G^k$, and the obtained results are very similar (see Colavitti et al. 2009). Note that the surface gas density exponent, {\itshape k}, is equal to 1.5. This value is in agreement with the results of Kennicutt (1998) and with N-body simulation results by Gerritsen \& Icke (1997). 
The efficiency of the SFR is set to $\tilde{\nu}=2$~Gyr$^{-1}$ during the halo-thick disk formation, while it is equal to 1~Gyr$^{-1}$ during the thin-disk formation, to ensure the best fit to the observational features in the solar vicinity, and becomes zero when the gas surface density drops below a certain critical threshold (Kennicutt 2001). The adopted threshold gas densities in the halo-thick disk and thin-disk phases are different. The physical reason for a threshold in the star formation is related to the gravitational stability, according to which, below a critical density, the gas is stable against density condensations and consequently the star formation is suppressed, the actual critical value depending also on the rotational properties of the galaxy. 
However, it is worth noting that the existence of such a  threshold has been challenged by some observational works (e.g. Leroy et al. 2008; Goddard et al. 2010) on the basis of the existence of star formation at large galactocentric distances found in some disks. Besides that, it is not clear how such a threshold should change according to the environment. Theoretical arguments (e.g. Elmegreen 1999) have suggested that a gas density threshold for star formation should exist also in objects suffering bursts of star formation but it should be lower than in disks. For this reason Chiappini et al. (2001) had chosen a threshold of 4~M$_{\odot}$ pc$^{-2}$ for the halo where the SF was stronger than in the thin disk. Here (see Section~4) we assume that the threshold in the thick disk is intermediate between that of the halo and that of the thin disk.

The IMF is that of Scalo (1986), assumed to be constant during the evolution of the Galaxy (see Chiappini, Matteucci \& Padoan 2000). The nucleosynthesis prescriptions are the same as in model~15 of Romano et al. (2010), as described in Section~4.

\section[]{The Three Infall Model} \label{three_infall_model}

We present now our new model for the chemical evolution of the Milky Way: the Three Infall Model (3IM). 
In this case, we have assumed three main infall episodes: the first is responsible for the creation of the halo, the second produces the thick-disk component and the third gives rise to the thin disk. As in the previous model, the material accreted by the Galactic disk comes mainly from extragalactic sources, and the Galactic disk is approximated by several independent rings, 2 kpc wide, without exchange of matter between them, and the basic equation is the same seen before, i.e. equation~(\ref{eq_modello}).

As this model assumes three main infall episodes, we had to change the rate of mass accretion in each shell:

\begin{displaymath}
\frac{d\sigma_i(r,t)_{inf}}{dt}= 
A(r)(X_i)_{inf}e^{-\frac{t}{\tau_H}}+B(r)(X_i)_{inf}e^{-\frac{t-t_{max_H}}{\tau_T}} 
\end{displaymath}
\begin{equation}
+C(r)(X_i)_{inf}e^{-\frac{t-t_{max_T}}{\tau_D}}
\label{rate_3im}
\end{equation}

where:

\begin{itemize}
\item $t_{max_H}$ is the time of maximum accretion onto the halo and roughly corresponds to the end of the halo phase,
\item $t_{max_T}$ represents the time of maximum accretion onto the thick disk, and roughly corresponds to the end of the thick-disk phase,
\item $\tau_H$ is the timescale for mass accretion in the halo,
\item $\tau_T$ is the timescale for mass accretion in the thick-disk component,
\item $\tau_D$ is the timescale for mass accretion in the thin-disk component.
\end{itemize}

Again, the quantities $A(r)$, $B(r)$, $C(r)$ are derived from the condition of reproducing the current total surface mass density distribution (halo+thick+thin disk), taken from Kuijken \& Gilmore (1991).

For the SFR, we have assumed the same functional form of the 2IM, in order to make a comparison between the results of the two models. The novelty introduced in the 3IM concerns the efficiency for the SFR, $\tilde{\nu}$. In this case, we have set $\tilde{\nu}=$~2 Gyr$^{-1}$ for the halo phase, $\tilde{\nu}=$~10 Gyr$^{-1}$ for the thick disk and $\tilde{\nu}=$~1 Gyr$^{-1}$ for the thin disk. This is to ensure the best fit to the observational features in the solar vicinity, as we will see in Section~5. Moreover, we have introduced a threshold in the SFR also in the thick-disk component, in addition to that for the halo and that for the thin disk. 
In Fig. 1, left panel, we show the SFR versus time as predicted by the 3IM for the halo, thick and thin disk. In Fig. 1, right panel, we show the SFR density, $\Sigma_{\mathrm{SFR}}$, as a function of the gas density, $\Sigma_{\mathrm{gas}}$, for the three components: at a given value of $\Sigma_{\mathrm{gas}}$, $\Sigma_{\mathrm{SFR}}$ can be higher or lower depending on the assumed value of $\tilde{\nu}$, namely, the efficiency of star formation. In our model, the highest $\Sigma_{\mathrm{SFR}}$ is observed in the thick-disk phase. 
It is worth noticing at this point that the choice of three different values for the star formation efficiency is not as arbitrary as it could seem at first glance. While in our approach the adopted value of $\tilde{\nu}$ for each Galactic component simply stems from the comparison of the model results with the observations, there is some empirical justification for a varying star formation efficiency depending on the environment. Indeed, observations have shown that the SFR is enhanced when two galaxies interact or coalesce (e.g. Patton et al. 2013, and references therein). If a gas-rich merger is the most important driver of thick-disk formation (Dierickx et al. 2010), our need for a more efficient star formation during the thick-disk phase may be related to a gas-rich merger origin for the Galactic thick disk\footnote{Mergers are not in contradiction with our assumption of formation from `infalling gas', as long as the accreted subunits are mostly gaseous.}.

In summary, our best 3IM assumes:
\begin{enumerate}
\item The total halo surface mass density profile is constant and corresponds to $\sigma_H(r,t_G)=$~17 M$_{\odot}$ pc$^{-2}$ at all Galactocentric distances;
\item The total thick-disk surface mass density profile is constant and corresponds to $\sigma_T(r,t_G)=$~24 M$_{\odot}$ pc$^{-2}$ (Chabrier 2003);
\item The total thin-disk surface mass density profile is exponential and corresponds to $\sigma_D(r_{\odot},t_G)=$~30 M$_{\odot}$ pc$^{-2}$, at the solar ring. It is worth noting that the sum of the values of $\sigma_T(r_{\odot},t_G)$ and $\sigma_D(r_{\odot},t_G)$ gives as result 54~M$_{\odot}$ pc$^{-2}$, which corresponds to the total surface mass density of the disk at the solar ring, as given by Kuijken \& Gilmore (1991).
\item $t_{max_H}=$~0.4 Gyr;
\item $t_{max_T}=$~2.0 Gyr; 
\item $t_G$ is set to be equal to 13.7~Gyr;
\item $\tau_H=$~0.2 Gyr;
\item $\tau_T=$~1.25 Gyr;
\item $\tau_D=$~6.0 Gyr at the solar radius;
\item The thresholds in the gas density are assumed to be 4, 5 and 7~M$_{\odot}$ pc$^{-2}$ during the halo, thick- and thin-disk phases, respectively (see discussion before).
\end{enumerate}

\begin{figure*}
\begin{center}
\begin{tabular}{cc}
\psfig{figure=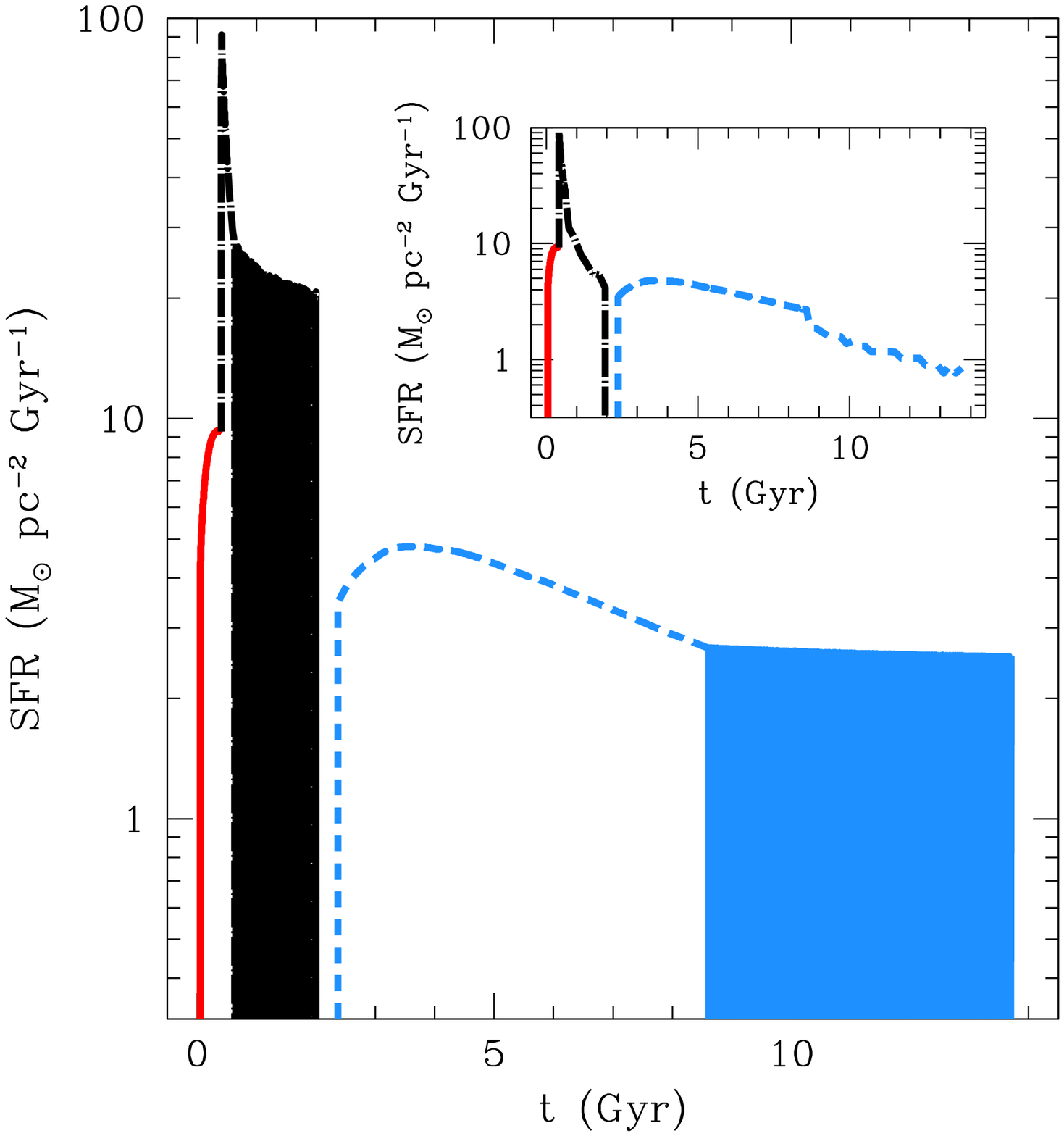,width=\columnwidth}
\psfig{figure=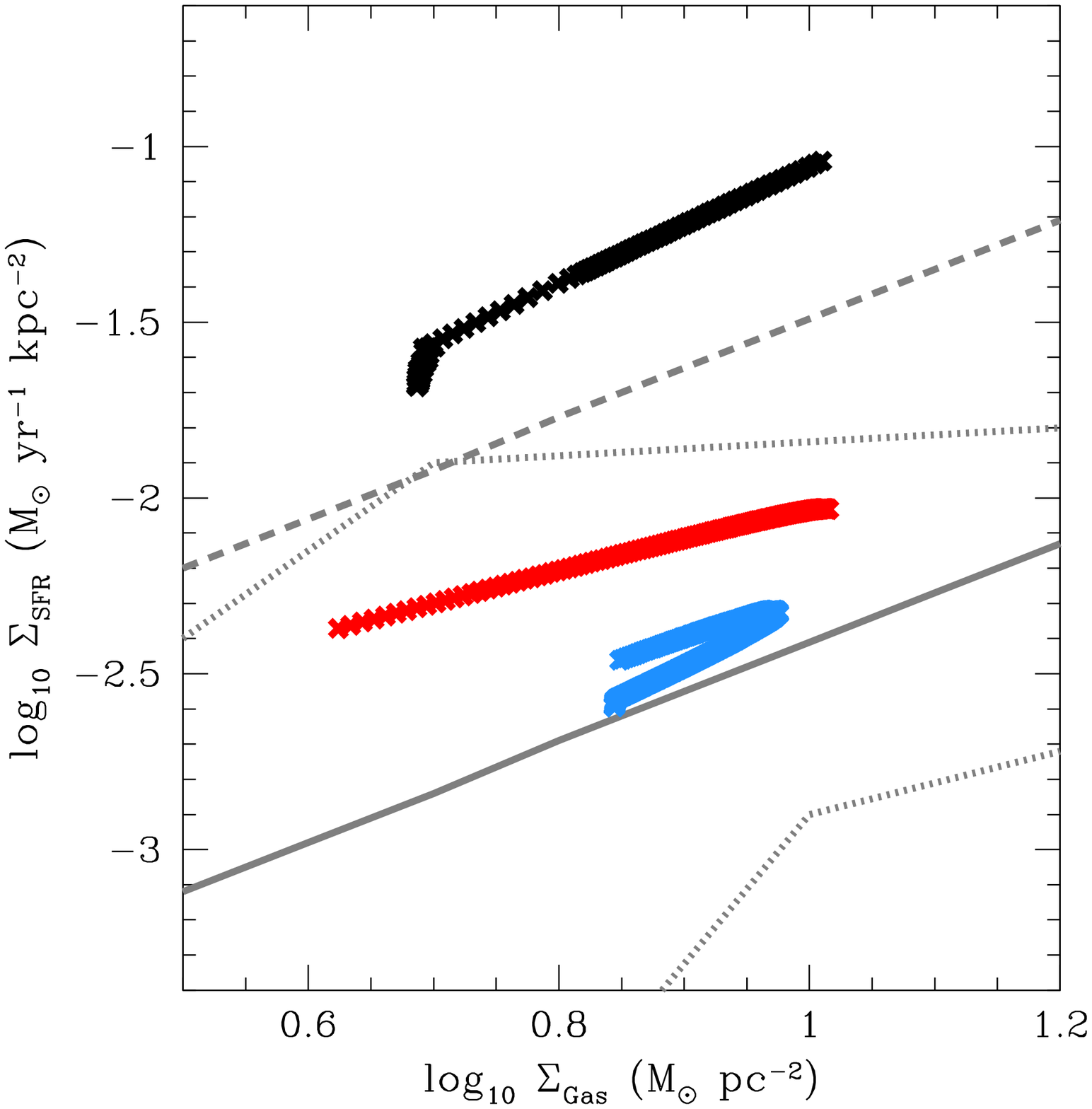,width=\columnwidth}\\
\end{tabular}
\caption{ Left panel: temporal evolution of the SFR as predicted by the 3IM. The (red) solid portion of the curve refers to the halo phase, the (black) dot-dashed one to the thick-disk phase, the (blue) dashed one to the thin-disk phase. In the inset, we provide the SFR averaged over timesteps of 200 Myr for phases in which the gas rapidly oscillates around the threshold value. Right panel: SFR density versus gas density in the halo (red crosses), thick disk (black crosses) and thin disk (blue crosses). Also shown are the fit to the $\Sigma_{\mathrm{SFR}}$--$\Sigma_{\mathrm{gas}}$ relation for local spirals and $z =$ 1.5 BzK galaxies by Daddi et al. (2010; gray solid line), the extrapolation of the starburst sequence from the same authors (gray dashed line) and the region of the plot occupied by spiral galaxies data (delimited by the dotted gray lines; see figure~2 of Daddi et al. 2010).}
\label{sfr_three_infall} 
\end{center}
\end{figure*}

\subsection{Nucleosynthesis Prescriptions} \label{nucleo_pre}

One of the most important ingredients for chemical evolution models are the nucleosynthesis prescriptions and the implementation of the yields in the model. 
In both 2IM and 3IM, we have used the same nucleosynthesis prescriptions in order to make a comparison between the two models. The prescriptions are the same as model~15 of Romano et al. (2010), where a detailed description of the adopted yield sets can be found. Here we only recall a few basic facts.

As regards to the computation of the stellar yields, one has to distinguish between different mass ranges, as well as single stars versus binary systems:

\begin{itemize}
\item low- and intermediate-mass stars (0.8~M$_{\odot}$--8~M$_{\odot}$) which are divided in single stars and binary systems eventually exploding as Type Ia SNe;
\item massive stars ($M>$~8 M$_{\odot}$);
\item novae.
\end {itemize}
 
\subsection{Low- and intermediate-mass stars}

\paragraph*{Single stars} The single stars in this mass range contribute to the galactic enrichment through planetary nebula ejection and quiescent mass loss along the giant branches. They enrich the ISM mainly in He, C, and N. They can also produce interesting amounts of $^7$Li, Na and $s$-process elements. For these stars, which end their lives as white dwarfs, we adopt the prescription of Karakas (2010).

\paragraph*{Type Ia SNe} Type Ia SNe are thought to originate from carbon deflagration in C-O white dwarfs in binary systems. Type Ia SNe contribute a substantial amount of iron (0.6~M$_{\odot}$ per event) and non negligible quantities of Si and S. They also contribute to other elements, such as O, C, Ne, Ca, Mg and Mn, but in negligible amounts compared with the masses of such elements ejected by Type II SNe. The adopted nucleosynthesis prescriptions are from Iwamoto et al. (1999). 

\subsection{Massive stars}

Massive stars are the progenitors of Type II SNe. If the explosion energies are significantly higher than 10$^{51}$~ergs, hypernova events may occur. For this range of masses, we adopt up-to-date stellar evolution calculations by Kobayashi et al. (2006) for the following elements: Na, Mg, Al, Si, S, Ca, Sc, Ti, Cr, Mn, Co, Ni, Fe, Cu and Zn. As for He and CNO elements, we take into account the results of Geneva models for rotating massive stars (see Romano et al. 2010 for references).

\subsection{Novae}

As for nova systems, which we consider to be binary systems constituted by a white dwarf and a main sequence or red giant companion, we include the products of explosive nucleosynthesis computed by Jos\'{e} ed Hernanz (1998). Novae contribute mainly to the enrichment in CNO isotopes and, perhaps, $^{7}$Li (Romano et al. 1999; Romano \& Matteucci 2003).

\section{Results} \label{res}
A good model of chemical evolution of the Galaxy should reproduce a number of constraints that is larger than the number of free parameters. Our set of observational constraints includes:

\begin{itemize}
 \item Metallicity distributions of long-lived stars belonging to the thick- and thin-disk components;
 \item Solar abundances;
 \item Present-time SFR;
 \item Type Ia and type II supernova rates at the present time;
 \item Variation in the relative abundances of the most common chemical elements;
 \item Present-time stellar, gas and total surface mass densities.
\end{itemize}

In this section we show the results of our best model which has been selected after several numerical simulations in which we changed several parameters. Before selecting our best model, we run several numerical simulations by varying, one at the time, the most important parameters of the model. These are: the star formation efficiency, the time scale for gas accretion and the threshold gas density.

\subsection{The Star Formation History}

\begin{figure}
%\begin{center}
\psfig{figure=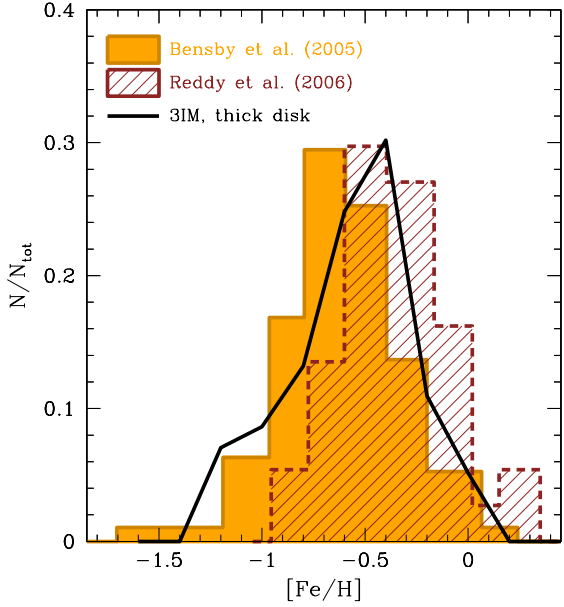,width=\columnwidth}
\caption{Theoretical metallicity distribution function from the Three Infall Model for the thick disk (solid line), compared with observations. Data are taken from Reddy et al. (2006; shaded histogram) and from Bensby et al. (2005; filled histogram).}
\label{mdf_thick_three_infall} 
%\end{center}
\end{figure}

As already mentioned, in Fig.~1 we show the SFR versus cosmic time (left panel) and gas density (right panel), as predicted by the 3IM. 
We recall that in  the 3IM we adopt three different gas thresholds for star formation, as described in the previous section. 
First of all, in this plot we see that, during the halo formation, the gas density never goes below the threshold, at variance with the 2IM, in which the SFR during the halo formation was oscillating because of the gas density threshold (see Chiappini et al. 1997, 2001). 
In this case, this occurs because in the 3IM we assume a time of maximum accretion for the halo, namely the time at which the halo formation ends, of 0.4~Gyr. This time is shorter than assumed in the 2IM and the infall rate is stronger, therefore,  it is more difficult to go below the threshold gas density of  4~M$_{\odot}$ pc$^{-2}$ during this time interval. 
Then, the thick-disk formation starts and the star formation efficiency $\tilde{\nu}$ changes from 2~Gyr$^{-1}$ to 10~Gyr$^{-1}$. We had to change the star formation efficiency in order to reproduce the metallicity distribution function and the trends of [O/Fe] and [$\alpha$/Fe] ratios versus [Fe/H] in the thick disk. Indeed, we also ran the model without changing the star formation efficiency, but the results were not in agreement with the observations, despite we had inserted a new episode of infall of primordial gas. Thus, in our opinion, the only way in which the thick disk can be reproduced, is that of setting a star formation efficiency an order of magnitude larger than the thin-disk one.

This change of $\tilde{\nu}$ is responsible for the abrupt increase in the SFR, from $\sim$10~M$_{\odot}$ pc$^{-2}$ Gyr$^{-1}$ to 90~M$_{\odot}$ pc$^{-2}$ Gyr$^{-1}$ seen in Fig. 1. After this spike the star formation trend presents an intermittent behavior regulated by the surface gas density in the thick disk. 
This intermittent behavior ends at 2 Gyr, which is set to be the time of maximum accretion of gas onto the thick disk. At this time, we find a star formation gap (like the gap in the 2IM between the end of the halo-thick disk and the beginning of the thin-disk phase), which lasts for 0.4~Gyr. The star formation efficiency $\tilde{\nu}$ changes again to the value of 1~Gyr$^{-1}$, which is appropriate for the thin disk. At this point, the formation of the thin disk starts and in this case the threshold of 7~M$_{\odot}$ pc$^{-2}$ is reached roughly at 8.5 Gyr, thus we find oscillations in the star formation from that epoch until the present time.

The SFR densities required to fit the Milky Way data in the framework of our model (Fig.~1, right panel, crosses) are compared to the fit to the $\Sigma_{\mathrm{SFR}}$--$\Sigma_{\mathrm{gas}}$ relation for local spirals and $z =$ 1.5 BzK galaxies by Daddi et al. (2010; Fig.~1, right panel, gray solid line), to the extrapolation of the Daddi et al. (2010) starburst sequence (Fig.~1, right panel, gray dashed line) and to the locus of spiral galaxies data (Daddi et al. 2010, and references therein; Fig.~1, right panel, gray dotted lines). While during the thin-disk phase our SFR densities are consistent with Daddi et al.'s (2010) relationship for normal spirals, during the thick-disk phase they are significantly higher and more consistent with the low-$\Sigma_{\mathrm{gas}}$ prosecution of the Sequence of Starbursts identified by Daddi et al. (2010, see also their figure~2). If the SFR is enhanced during mergers, we are forced to conclude that a gas-rich merger origin of the Galactic thick disk is favoured on the basis of chemical arguments.

\begin{figure}
%\begin{center}
\psfig{figure=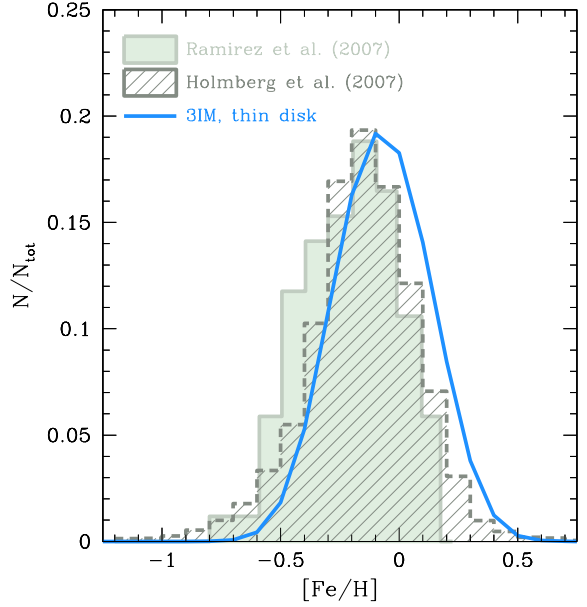,width=\columnwidth}
\caption{Theoretical metallicity distribution function from the Three Infall Model for the thin disk (blue solid line), compared with observations. Data are taken from Ram{\'{\i}}rez et al. (2007; filled histogram) and from the Geneva-Copenhagen Survey (GCS; Holmberg, Nordstr\"om and Andersen 2007; shaded histogram).}
\label{mdf_thin_three_infall} 
%\end{center}
\end{figure}

\subsection{Metallicity Distribution Function}

The first observational constraint which we have to analyze is the MDF of dwarf stars both in the thick and thin disk. 
Because of the introduction of a new episode of infall of gas, which gives rise to the thick disk, now we are able to create two MDFs using the 3IM: one for the thick disk and one for the thin disk; they are visible in Figs. 2 and 3, respectively. 
Let us consider the first plot (Fig. 2): it is clear from the data (filled and shaded histograms in Fig.~2) that the metallicity of the thick disk extends from [Fe/H]$\sim-$1.3 dex to [Fe/H]$\sim$ 0~dex and the mean value is $<$[Fe/H]$>\sim-$0.5 dex. 
Our best model (solid line in Fig. 2) well reproduces the observations. 
The relative number of stars with [Fe/H]$\sim-$0.5 predicted by the model is equal to $\sim 30\%$. Then we have a tail of metal-poor stars equal to $\sim 2\%-8\%$ and the same relative number for higher-metallicity stars. 
Therefore, concerning the thick disk, we believe that our model is good from the chemical point of view, as it predicts a relative number of stars in agreement with the observations, in the right [Fe/H] intervals.

In Fig.~3, we can see that the MDF of the thin disk extends from [Fe/H]$\sim-$0.5~dex to [Fe/H]$\sim +$0.3~dex and the mean value is $<$[Fe/H]$>\sim-$0.1~dex. 
Also for the thin disk, the agreement between model predictions (solid line) and data (filled and shaded histograms) is good. Perhaps, there is a little displacement with respect to Ram{\'{\i}}rez et al.'s (2007) data (filled histogram), due to the adopted selection criteria. In fact, in Ram{\'{\i}}rez et al. (2007) different Galactic space velocities ($U_{LSR}, V_{LSR}$ and $W_{LSR}$) of the stellar populations were used. This can imply a different sample of thin-disk stars. In particular, thick-disk stars might have been considered as thin-disk members because of their higher Galactic space velocities. This explains why there are many members with [Fe/H]$\sim-$0.5~dex in the Ram{\'{\i}}rez et al. (2007) MDF. 
The predicted relative number of stars with [Fe/H]$\leq-$0.4~dex is $\sim 2\%$, while at high metallicities, especially around [Fe/H]$\sim$0.2, the relative number of stars become $\sim 18\%$. 
It is worth noting that our best 3IM reproduces very well the high metallicity tail of the G-dwarf metallicity distribution, whereas the 2IM was lacking of the most metal-rich thin-disk stars. Therefore, with our model there is no need to invoke stellar migration fron the inner disk regions to explain the high-metallicity tail of the observed distribution.

\subsection{Solar Abundances}

The solar abundances predicted by the 3IM are compared with observations (Asplund et al. 2009), as well as predictions from the latest 2IM and CMG97 in Table \ref{solar_three_infall}. The abundances are expressed as 12+log(X/H) where X is the abundance by number of a generic chemical element and 12 is the weight attributed to hydrogen. 
These abundances represent the composition of the ISM at the time of the formation of the Sun, 4.5 Gyr ago. Since we assume a Galactic lifetime of 13.7 Gyr, we have computed the solar abundances at 9.2 Gyr after the Big Bang. 
Given the uncertainties involved in the observed solar abundances as well as in the theoretical yields, we can consider that the model is in quite good agreement with the observed value within a factor 0.5 of difference. 
As we can see from Table 1, the solar abundances of the 3IM are very similar to those predicted by the 2IM and to the observed ones. 
This suggests that the insertion of the thick-disk component has not altered theresults of the 2IM for the thin disk. 
The only abundance which is in disagreement with the data is that of magnesium, but this is a well known problem, that is due to the too low magnesium yields from Type II supernovae predicted in most nucleosynthesis computations.  

\begin{table}
\begin{center}
\begin{tabular}{|c|c|c|c|c|}
\hline
Elem. & Observations & 2IM & CMG97 & 3IM\\
\hline
O & $8.69\pm0.05$ & 8.78 & 8.71 & 8.81\\
Mg & $7.60\pm0.04$ & 7.37 & 7.21 & 7.40\\
Si & $7.51\pm0.03$ & 7.59 & 7.66 & 7.64\\
Ca & $6.34\pm0.04$ & 6.22 & 6.23 & 6.27\\
Fe & $7.50\pm0.04$ & 7.56 & 7.48 & 7.64\\
\hline
\end{tabular}
\end{center}
\caption{Solar Abundances. Model abundances evaluated at 9.2 Gyr, compared with observations (Asplund et al. 2009), CMG97 and 2IM.}
\label{solar_three_infall}
\end{table}

\begin{figure}
\begin{center}
\psfig{figure=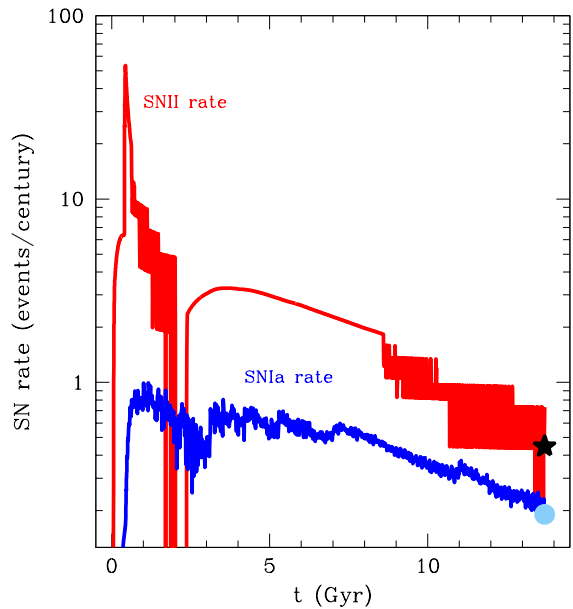,width=\columnwidth}
\caption{Temporal evolution of Type Ia (lower blue line) and Type II (upper red line) SN rates, as predicted by the Three Infall Model. The black star and the light-blue filled circle are the observed values at the present time after Li et al. (2011).}
\label{sn_rate_three_infall} 
\end{center}
\end{figure}

\subsection{Supernova Rates}

To compare with the predictions of the 3IM we used the SN rates in the Galactic disk estimated by Li et al. (2011), both for Type Ia SNe and Type II SNe.

The observed present-time SN rates are:
\begin{itemize}
\item (0.189 $\pm$ 0.031) SNe century$^{-1}$ for Type Ia SNe;
\item (0.462 $\pm$ 0.044) SNe century$^{-1}$ for Type II SNe;
\end{itemize}
while the rates predicted by the 3IM are:
\begin{itemize}
\item 0.189 SNe century$^{-1}$ for Type Ia SNe;
\item 0.456 SNe century$^{-1}$ for Type II SNe;
\end{itemize}
the agreement between model predictions and observations is excellent. 
As we can see in Fig. 4, the gap between the end of the thick disk and the beginning of the thin disk due to the adopted gas density threshold is present in the predicted SNII rate as well; this gap was present also in the 2IM predictions. The duration of the gap is 0.4~Gyr. The SN II rate shows a peak around $\sim$0.4 Gyr, which roughly corresponds to the end of the halo phase, and then it reaches the value of 50~SNe century$^{-1}$ because of the increased star formation efficiency during the thick-disk formation. 
After this spike, the SN II rate goes to zero at $\sim$~2 Gyr after the Big Bang, this time corresponding to the end of the thick disk and the beginning of the gap in the SFR. 
Once the thick-disk formation ends, star formation starts again and the number of supernovae per century increases until 4 Gyr, and then it decreases until the achievement of the present-time rate. 
On the other hand, Type Ia SNe, which are produced by progenitors with long lifetimes, show a smaller effect caused by the threshold in the star formation and, as shown in Fig. \ref{sn_rate_three_infall}, their rate first increases with time, then, it remains almost constant until it decreases and reaches the present-time value.

\begin{figure}
%\begin{center}
\psfig{figure=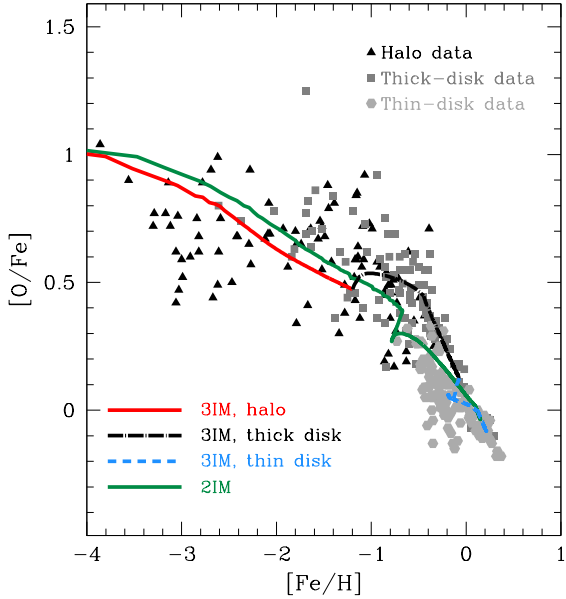,width=\columnwidth}
\caption{ [O/Fe] versus [Fe/H] behavior for the 3IM (red solid, black dot-dashed and blue dashed line) and the 2IM (green solid line). Halo data (black triangles) are taken from Cayrel et al. (2004) and Akerman et al. (2004). Thick-disk data (dark-gray squares) are taken from Gratton et al. (2003), Bensby et al. (2005) and Ramya et al. (2012). Thin-disk data (light-gray hexagons) are taken from Reddy et al. (2003) and Bensby et al. (2005).}
\label{[O/Fe] vs [Fe/H]_three} 
%\end{center}
\end{figure}

\subsection{Relative Abundances}

The last observational constraints considered are the [X/Fe] versus [Fe/H] relations. As mentioned before, the abundance ratios are less dependent on model parameters than the absolute abundances, since they depend essentially on the nucleosynthetic yields and on the IMF. In the following, we show and explain the behavior of the relative abundances of oxygen and three $\alpha$-elements, i.e. magnesium, silicon and calcium, with respect to iron as functions of [Fe/H].

\subsubsection{The [O/Fe] versus [Fe/H] relation}

Fig.~\ref{[O/Fe] vs [Fe/H]_three} shows the oxygen-to-iron ratio with respect to iron relative to the Sun. Our best model is represented by the multi-coloured line in this figure: the red (solid) portion of the curve is for the halo phase, the black (dot-dashed) one is for the thick disk and the blue (dashed) one is for the thin disk. The predictions of the 2IM are also reported for comparison (green solid line). 
We immediately see that the agreement between model predictions and data is improved by the introduction of a new episode of infall of gas responsible for the formation of the thick disk: indeed, the 3IM fits very well the thick-disk data (dark-gray squares). The trend can be explained through the different roles played by the two types of SNe involved in the Galactic abundance enrichment. In fact, while oxygen is produced only by Type II SNe, which have high-mass and short-lifetime progenitors, iron is mainly produced on longer timescales by Type Ia SNe, which are believed to be the result of the explosion of C-O white dwarfs in  binary systems.

The halo trend of the oxygen-to-iron ratio is similar to that of the 2IM, as expected. Then the abundance ratio decreases until [Fe/H]$\sim-$1.2~dex , roughly 1.0~Gyr after the Big Bang. At this point, using the 3IM, we find a new feature due to the new episode of infall of gas. In fact, during this infall, we suppose that the star formation efficiency increases from 2~Gyr$^{-1}$ to 10~Gyr$^{-1}$, therefore there is a higher number of Type II supernovae which contributes to the oxygen overabundance. Thus, instead of finding the knee, we predict a bump, which is present also in the thick-disk data. 
Then, the [O/Fe] ratio slightly decreases until [Fe/H]$\sim-$0.2~dex, which corresponds to the end of the formation of the thick disk, and the star formation efficiency changes to 1~Gyr$^{-1}$ until the end of the formation of the thin disk, which occurs with a timescale of $\sim$6.0 Gyr. The reason for the decline in the predicted [O/Fe] with increasing metallicities is obviously  due to the explosion of Type Ia supernovae, which are the main producers of iron (time-delay model). 
In reality, the bump at the end of the halo phase and the knee at the end of the thick-disk phase are due not only to the change of the star formation efficiency, but also to the adopted gas density thresholds for star formation. These thresholds are, in fact, responsible for the strong dilution of the ISM abundances at the moment of maximum infall onto the disk and, coupled with the different SFRs, they produce the discontinuity seen in Fig. \ref{[O/Fe] vs [Fe/H]_three}. 
We also ran the 3IM without any gas threshold. The result (not shown in the figures) was a displacement in the two MDFs towards higher metallicities and a worse fit as regards to the [$\alpha$/Fe] versus [Fe/H] relations in the thick disk. Thus, we conclude that, in the framework of the model presented here, the thresholds are necessary in order to reproduce the relevant data. 
Finally, in Fig. 5 it is evident that the gap in the SFR occurs before that in the 2IM. This is due again to the faster evolution and consequent faster increase of the Fe abundance during the halo phase predicted by the 3IM.

\begin{center}
\begin{figure*}
\begin{tabular}{cc}
\psfig{figure=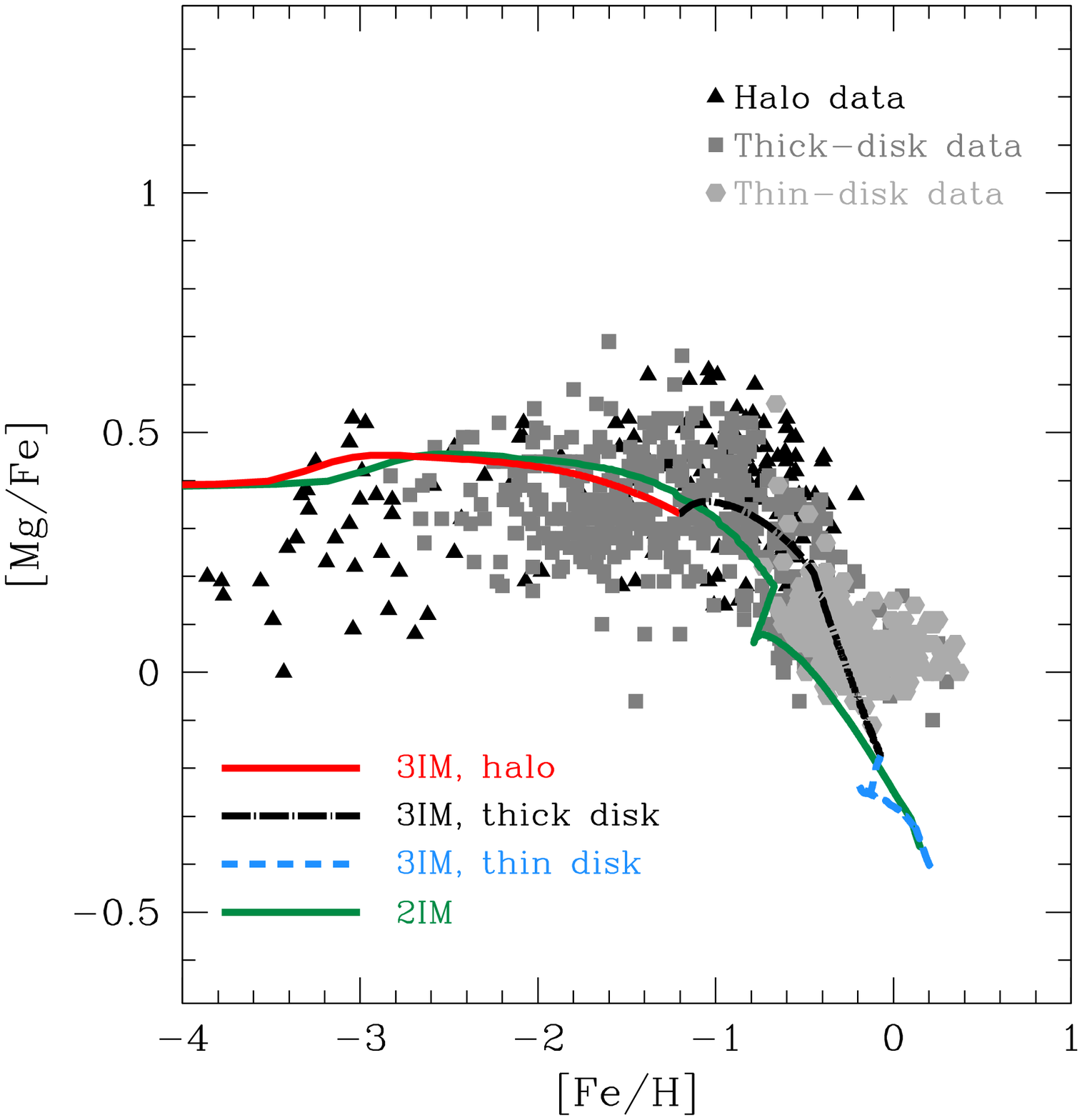,width=\columnwidth}
\psfig{figure=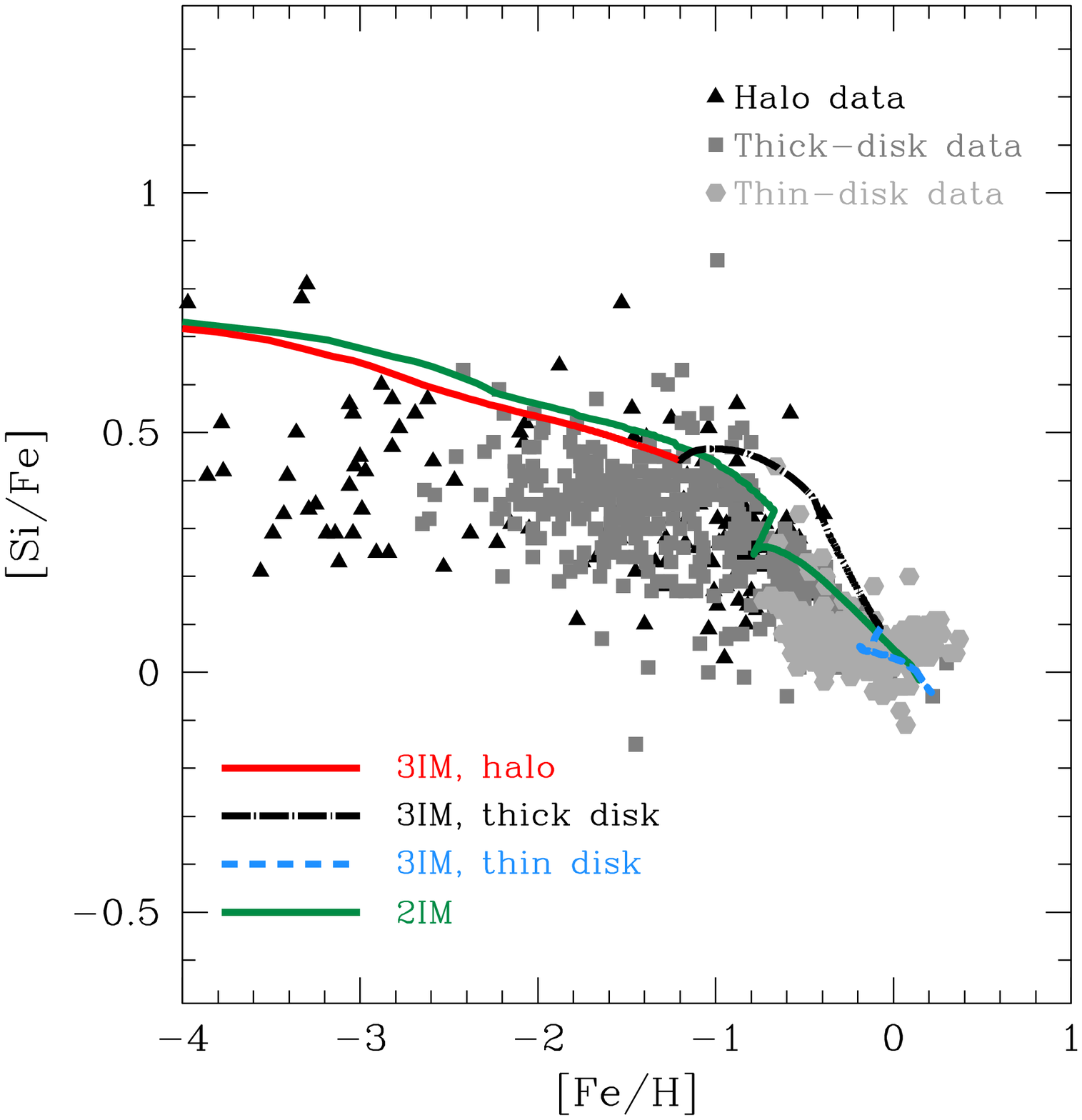,width=\columnwidth}\\
\end{tabular}
\caption{Predicted behavior of the relative ratios  of Mg (left panel) and Si (right panel) to iron as a function of the relative iron abundance, for the 3IM (red solid, black dot-dashed and blue dashed lines) and the 2IM (green solid lines). Halo data (black triangles) are taken from Gratton et al. (2003), Cayrel et al. (2004), Reddy et al. (2006), Mashonkina et al. (2007) and Shi et al. (2009). Thick-disk data (dark-gray triangles) are taken from Gratton et al. (2003), Reddy et al. (2006), Bensby et al. (2005), Ruchti et al. (2011) and Ramya et al. (2012). Thin-disk data (light-gray hexagons) are taken from Gratton et al. (2003), Reddy et al. (2003), Reddy et al. (2006) and Bensby et al. (2005).}
\label{alpha_elements_three_infall}
\end{figure*}
\end{center}

\subsubsection{The [X/Fe] versus [Fe/H] relation for the other $\alpha$-elements}

\begin{center}
\begin{figure}
\begin{tabular}{cc}
\psfig{figure=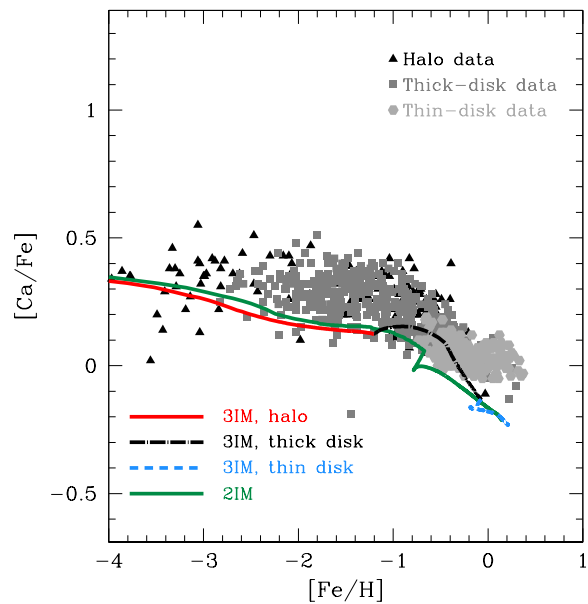,width=\columnwidth}
\end{tabular}
\caption{Predicted behavior of [Ca/Fe] versus [Fe/H] for the 3IM (red solid, black dot-dashed and blue dashed line) and the 2IM (green solid line). Halo data (black triangles) are taken from Gratton et al. (2003), Cayrel et al. (2004), Reddy et al. (2006) and Mashonkina et al. (2007). Thick-disk data (dark-gray squares) are taken from Gratton et al. (2003), Reddy et al. (2006), Bensby et al. (2005), Ruchti et al. (2011) and Ramya et al. (2012). Thin-disk data (light-gray hexagons) are taken from Gratton et al. (2003), Reddy et al. (2003), Reddy et al. (2006) and Bensby et al. (2005).}
\label{alpha_elements2_three_infall}
\end{figure}
\end{center}

Figs.~\ref{alpha_elements_three_infall} and \ref{alpha_elements2_three_infall} show the predicted abundances of $\alpha$-elements (Mg, Si and Ca) relative to iron as functions of [Fe/H], compared with the observational data. 
The new model predicts overabundances of $\alpha$-elements relative to iron almost constant until [Fe/H]$<-$1.2~dex. Then, as seen before with the oxygen behavior, the trend shows a bump (again due to the coupled effect of the change in the star formation efficiency and the thresholds), which is followed by a slight decline until [Fe/H]$<-$0.2~dex, due to the delayed explosion of Type Ia SNe. 
The observed abundance trends are well defined and smooth, with a small overall scatter. This is indicative of the fact that both the thin and the thick disks formed from interstellar gas that was reasonably well mixed. 
The star formation in the thick disk must have been fast initially, to allow the formation of stars with high [$\alpha$/Fe] ratios before the significant enrichment from Type Ia SNe at higher metallicities, although the Type Ia SN signature is already evident, as shown by Chiappini (2008). 
Finally, we enter in the metallicity range characteristic of the thin-disk component, in which the [$\alpha$/Fe] ratios decrease until [Fe/H]$\sim$0.3~dex, due to the increasing contribution of SNeIa to Fe production. 
We notice that the spread in the thin-disk data is due mainly to the fact that the data originate from different sources. The agreement between model predictions and observations is improved by the introduction of the new episode of infall of gas, responsible for the formation of the thick disk. We conclude that the agreement between predictions and observations is generally satisfactory (though Ca and Si are slightly under- and over-produced, respectively, over the full metallicity range because of uncertienties in their yields), except for Mg at the highest thin-disk metallicities. However, we note that the steep decline of [Mg/Fe] at the highest metallicities is due to the adopted yields, that are known to underestimate the Mg production from massive stars (see, e.g., Kobayashi et al. 2006; Romano et al. 2010). Moreover, the observational data show that [Mg/Fe] flattens considerably at high metallicity at variance with what happens for O, while Si and Ca seem to show also some flattening. This flattening, not explained by the current available yields, could perhaps be explained by a larger Mg production from Type Ia SNe, or by invoking larger Mg yields from SNeII at supersolar metallicities. The fact that bulge stars at similar, and even higher, metallicities display exactly the same flattening in the [$\alpha$/Fe] ratios, for the same $\alpha$s but oxygen (Bensby et al. 2013), strongly point to the need to use different (higher) SNII yields for $\alpha$-elements for metallicities higher than solar.

\section{Discussion and Conclusions} \label{disc_conc}

Many different scenarios have been proposed for the formation of the thick-disk component of the Galaxy, among which heating by mergers (Quinn et al. 1993; Villalobos et al. 2010), accretion of satellites (Abadi et al. 2003), early and clumpy rapid star formation (Bournaud et al. 2009), radial stellar migration (Loebman et al. 2011; Curir et al. 2012) and combination of these (Minchev, Chiappini \& Martig 2012). Likely, all these processes have acted in the Milky Way to some extent. Yet, it is not clear which is the dominant one.
 
In this paper, we present a model which can explain the chemical evolution of the halo, thick and thin disk of the Milky Way. 
The underlying idea is the following: the bulk of the stars that are found in these three components results from in situ star formation triggered by three main infall episodes. 
We explored several cases in which we varied in turn the gas density threshold, the star formation efficiency and the timescale for the formation of each Galactic component trying to find a model in agreement with all the observational constraints. This procedure has allowed us to determine values of the parameters which bring the theoretical predictions in agreement with the observations.

Our best model suggests that during the first infall of primordial gas, which lasts approximatively 0.4~Gyr, the halo rapidly forms with a star formation efficiency of 2~Gyr$^{-1}$. 
Once the halo is created, a slower infall of primordial gas, which lasts $\sim$2~Gyr, gives rise to the thick disk. During the thick-disk phase, the star formation efficiency changes from 2~Gyr$^{-1}$ to 10~Gyr$^{-1}$. We have to assume such a high star formation efficiency in order to reproduce two fundamental observational constraints, namely the MDF and the [$\alpha$/Fe] ratios of thick-disk stars. We have also ran the model without changing the star formation efficiency, but the result was to worsen the very good agreement between model predictions and data. An efficiency of  10~Gyr$^{-1}$ indicates that  the thick disk formed in a starburst. A similar suggestion was put forward by Kroupa (2002), who claimed that the thick disk formed via an early, strong star formation activity, leading to the formation of large stellar clusters that expelled their stars thus forming the thick-disk component. Hubble Ultra Deep Field observations showing thick disks in formation also suggest that the thick disks form during a highly turbulent gas phase, before the accreted material cools down to give rise to the thin disks (Elmegreen \& Elmegreen 2006). 

The third infall episode is responsible for the formation of the thin disk. It has a much slower timescale for mass accretion ($\sim$6.0~Gyr) and a much lower star formation efficiency (1~Gyr$^{-1}$), values in agreement with our previous 2IM. 
In the thin disk, the star formation is actually still active and we predict an intermittent behavior, regulated by the threshold in the surface gas density which is assumed to be 7~M$_{\odot}$ pc$^{-2}$, a result already found and discussed in CMG97. The most important achievement of this scenario, which assumes three main, independent episodes of gas accretion during the Galaxy lifetime, is that it is possible to well reproduce the metallicity distributions of thick- and thin-disk stars in the solar vicinity, separately. In the 2IM model that was not possible since the thick disk was included in the halo formation. On the other hand, one of the main problems of models assuming that the thick disk formed from gas shed by the halo (e.g. Pardi et al. 1995) was the failure in reproducing the MDF of halo and thick-disk stars at the same time. This suggests that the thick disk should have formed independently from the halo. Moreover, the 3IM is able to fit well the abundance trends of oxygen and other $\alpha$-elements (magnesium, silicon and calcium) in the three Galactic components. These trends are explained by the delayed explosions of Type Ia supernovae with respect to the early explosion of Type II SNe, and by the more efficient SFR during the thick-disk phase. 
The very good agreement between model predictions and data is confirmed when we compute the solar abundances and present-time supernova rates. 
Though by no means unique, our three infall scenario offers a formation picture consistent with the majority of the observational constraints available for the Milky Way and suggests that gas accretion should be favored over accretion of stellar debris as the main formation mechanism of the thick disk. In particular, we find that the chemical properties of thick-disk stars can be reproduced by a model that envisages an independent episode of accretion of matter of primordial, or nearly primordial, chemical composition triggering an intense star formation activity. Independent observational evidence supports the gas-rich merger scenario as the most important mechanism of thick-disk formation (Dierickx et al. 2010; but see also Bournaud et al. 2009). Although gas accretion is crudely modelled in our model with a smooth exponential law fading in time, it is important to note that a merger origin for the thick disk is not ruled out, as long as the merging fragments are mostly gaseous. It is also important to notice that thick-disk stars define tight abundance trends, with spreads well within the observational errors characterizing the abundance determinations (e.g. Bensby et al. 2005). This is a clear indication that thick-disk stars must have formed in situ from a well homogenized medium. This makes the hypothesis of thick-disk formation through assembly of small stellar systems highly unlikely, since in this case we would expect much more loose abundance trends owing to the different evolutionary paths followed by the different merging structures.

Our main conclusions are as follows:

\begin{enumerate}
\item Our best model predicts stellar metallicity distributions in the thick and thin disks in very good agreement with the relevant observations (Reddy et al. 2006; Bensby et al. 2005; Ram{\'{\i}}rez et al. 2007; Holmberg et al. 2007). The MDF constitutes the most important constraint for chemical evolution models at present and implies a timescale for the thin-disk formation much longer than that for the formation of the halo and thick disk. The good fits to the MDFs of thick- and thin-disk stars suggest that the thick disk should have formed on a timescale of $\sim$1.25~Gyr, while the thin disk should have formed on a timescale of $\sim$6~Gyr in the solar vicinity. 
The simultaneous agreement between the stellar metallicity distribution of both thick and thin disk and the observational data in the solar neighborhood constitutes an important and new result, being a consequence of some main assumptions of our model. First, the decoupling between the rate of gas loss from the halo and the accretion of gas onto the thick and thin disk. This implies that the  accreted gas in all the infall episodes should be substantially unprocessed. 
Then another reason is represented by the adopted gas density threshold for star formation that limits the star formation in the thick-disk phase, allowing us to reproduce the peak of the MDF at [Fe/H]$\sim-$0.5 dex .  The threshold is not reached during the halo phase because of the short halo timescale (0.2 Gyr) and consequent fast evolution of this component, while it is active in the formation of the thick and thin disk. In the thin disk, such a threshold prevents the gas to attain a metallicity larger than [Fe/H]$\sim +$0.3 dex.
\item The predicted solar abundances are in very good agreement with those of Asplund et al. (2009), except for magnesium, because, for this chemical element, the yields from Type II SNe are too low (see Romano et al. 2010 for a discussion).
\item The present Type Ia and Type II SN rates result to be equal to 0.189 SNe century$^{-1}$ for Type Ia SNe and 0.456 SNe century$^{-1}$ for Type II SNe, in excellent agreement with the observed ones.
\item As regards to the abundance ratios of the four studied $\alpha$-elements (O, Mg, Si and Ca) relative to iron, our best model predicts an improved agreement with the data, due to the introduction of the new episode of infall of gas responsible for the formation of the thick disk. The only element for which the agreement is less good is Mg, for the reasons discussed above. In particular, for what concerns the predicted [Mg/Fe] ratio, there appears that it is too low at high [Fe/H] values. While the low predicted solar abundance could be solved just by increasing the Mg yields of massive stars, the flat observed [Mg/Fe] trend at higher metallicities is more difficult to interpret. In fact, according to the time-delay model which interprets the abundance ratios on the basis of the roles played by different SNe, a constant [X/Fe] ratio close to zero indicates that the element X is produced on the same timescales as the element Fe. So, perhaps more Mg should originate from Type Ia SNe at high metallicity, implying also a metallicity dependence for the Type Ia SN yields. 
The new important result of the 3IM is the fit to the thick-disk data, which shows that the thin- and thick-disk abundance trends are clearly separated, especially for the $\alpha$-elements (see Bensby et al. 2005). In particular, we find a new feature in the plot of [$\alpha$/Fe]: a bump in the  [$\alpha$/Fe] ratios which allows us to fit the thick-disk data. This bump is due to the high star formation efficiency assumed for the thick disk (10~Gyr$^{-1}$).
\item The assumption of a gas density threshold for star formation naturally leads to a star formation gap between the end of the thick-disk phase and the beginning of the thin-disk formation. This star formation gap is reflected in a more or less pronounced decrease in the [$\alpha$/Fe] ratios around [Fe/ H]$\sim-$0.2 dex, depending on the duration of the gap itself. In this work, we use a fixed value of the gas density threshold for each Galactic component: 4~M$_{\odot}$ pc$^{-2}$ for the halo, 5~M$_{\odot}$ pc$^{-2}$ for the thick disk and 7~M$_{\odot}$ pc$^{-2}$ for the thin disk. The existence of such thresholds is suggested by both observational and theoretical arguments (Kennicutt 1989; Elmegreen 1999; but see e.g. Leroy et al. 2008; Goddard et al. 2010). We suggest a duration of 0.4 Gyr for the gap.
\end{enumerate}
Our model can be further improved if we take into account the following points:
\begin{enumerate}
\item It is possible to use the 3IM to display radial profiles;
\item It is also possible to study other chemical elements, and to compare their abundance ratios with previous results;
\item Usually, in chemical evolution models, the star formation efficiency assumed for the bulge is 10--20~Gyr$^{-1}$ and the timescale for bulge formation is short (0.3--0.5 Gyr) (e.g. Ferreras et al. 2003; Ballero et al. 2007; Cescutti \& Matteucci 2011). Therefore, there is a certain similarity to the adopted star formation efficiency for the thick disk. Alves-Brito et al. (2010) and Mel{\'e}ndez et al. (2008) have pointed out the similarity in the abundance ratios of bulge and thick-disk stars. It is worth to explore better this similarity in the future. 
\item In order to impose constraints on the chemical evolution of spiral galaxies, the 3IM can be applied, with appropriate modifications, to other spirals.
\end{enumerate}

\section*{Acknowledgments}

We want to thank  M. Bergemann and P. Fran\c{c}ois for their very useful suggestions about the choice of the observational data. F.M. thanks Cristina Chiappini for many illuminating discussions. F.M. and D.R. acknowledge financial support from the project PRIN~MIUR~2010--2011 ``Chemical and Dynamical Evolution of the Milky Way and Local Group Galaxies'', prot.~2010LY5N2T. Finally, we would like to thank an anonymous referee for valuable suggestions which have improved the paper.

\label{lastpage}

\end{document}